\documentclass[prb,aps,amssymb,twocolumn,superscriptaddress,longbibliography]{revtex4-1}
\usepackage{amsmath}
\usepackage{amssymb}
\usepackage{amsthm}
\usepackage{amsfonts}
\usepackage{listings}

\usepackage{physics}
\usepackage{multirow}
\usepackage{diagbox}

\usepackage{enumerate}
\usepackage{latexsym}
\usepackage{color}
\usepackage{bm}
\usepackage{hyperref}
\hypersetup{
 pdfnewwindow=true, colorlinks=true,
 linkcolor=blue, anchorcolor=blue,
 citecolor=blue, filecolor=blue,
 menucolor=blue, urlcolor=blue}

\newcommand{\beginsupplement}{%
        \setcounter{table}{0}
        \renewcommand{\thetable}{S\arabic{table}}%
        \setcounter{figure}{0}
        \renewcommand{\thefigure}{S\arabic{figure}}%
     }

\usepackage{psfrag}

\usepackage{bm}
\usepackage{graphicx}
\usepackage{subfigure}

\RequirePackage[normalem]{ulem} 
\RequirePackage{color}\definecolor{RED}{rgb}{1,0,0}\definecolor{BLUE}{rgb}{0,0,1} 

\newcommand{\bsigma}{{\vb* \sigma}}

\newcommand{\bk}{{\vb* k}}

\newcommand{\bkp}{{\vb* k_{\parallel}}}
\newcommand{\bkpsq}{{\vb* k_\parallel^2}}

\newcommand{\I}{\mathrm{\uppercase\expandafter{\romannumeral1}}}
\newcommand{\II}{\mathrm{\uppercase\expandafter{\romannumeral2}}}
\newcommand{\III}{\mathrm{\uppercase\expandafter{\romannumeral3}}}
\newcommand{\IV}{\mathrm{\uppercase\expandafter{\romannumeral4}}}

\def\ie{{\it i.e.},\ }

\input{epsf}

\begin{document}

\tolerance 10000

\newcommand{\vk}{{\bf k}}

\draft

\title{Hinge Majorana Flat Band in Type-II Dirac Semimetals}


\author{Yue Xie}
\affiliation{Beijing National Laboratory for Condensed Matter Physics,
and Institute of Physics, Chinese Academy of Sciences, Beijing 100190, China}
\affiliation{University of Chinese Academy of Sciences, Beijing 100049, China}

\author{Xianxin Wu}
 \affiliation{CAS Key Laboratory of Theoretical Physics, Institute of Theoretical Physics,
Chinese Academy of Sciences, Beijing 100190, China}

\author{Zhong Fang}
\affiliation{Beijing National Laboratory for Condensed Matter Physics,
and Institute of Physics, Chinese Academy of Sciences, Beijing 100190, China}
\affiliation{University of Chinese Academy of Sciences, Beijing 100049, China}

\author{Zhijun Wang}
\email{wzj@iphy.ac.cn}
\affiliation{Beijing National Laboratory for Condensed Matter Physics,
and Institute of Physics, Chinese Academy of Sciences, Beijing 100190, China}
\affiliation{University of Chinese Academy of Sciences, Beijing 100049, China}


\begin{abstract}
Type-II Dirac semimetals exhibit a unique Fermi surface topology, which allows them to host novel topological superconductivity (TSC). We reveal a novel inter-orbital superconducting state, corresponding to the $B_{1u}$ and $B_{2u}$ pairings under the $D_{4h}$ point group. Intriguingly, we find that both first- and second-order TSC coexist in this novel state. It is induced by a dominant inter-orbital attraction and possesses surface helical Majorana cones and hinge Majorana flat bands, spanning the entire $z$-directed hinge Brillouin zone. Further investigation uncovers that these higher-order hinge modes are robust against the $C_{4z}$ symmetry-breaking perturbation.
\end{abstract}

\maketitle

\paragraph*{Introduction.---}
Majorana fermions, quasiparticle excitations in topological superconductors, are known for their non-Abelian statistics~\cite{NA1,NA2,NA3,NA4,NA5} and potential application in fault-tolerant quantum computing~\cite{FT1,FT2,FT3,FT4,FT5}. Their unique properties have generated significant interest in realizing Majoranas in various topological materials~\cite{Fu_Kane,Alicea,Qi_Wen,Qi,SrRuO,WS1,WS2,TSC_6}, including Dirac semimetals (DSMs)~\cite{Sato_TCSC,Node_vortex,HOTDS}. Recently, numerous $D_{4h}$-symmetric type-II DSMs have been reported in theories and experiments, such as VAl$_3$~\cite{VAl3,expVAl3}, KMgBi~\cite{KMgBi,expKMgBi}, and full Heusler compounds HfInPd$_2$ and YPd$_2$Sn~\cite{Heusler,HfInPd2,YPd2Sn}. Furthermore, some of them are found to exhibit intrinsic superconductivity at low temperatures~\cite{HeuslerSC1,HeuslerSC2,HeuslerSC3,KMgBiSC}. As the Lorentz-invariance-violating velocity tilts the Dirac cone from type I to type II, the Fermi surface (FS) experiences a Lifshitz transition from a closed sphere to an open pocket~\cite{Type_II}. This transition provides a distinct Fermi surface and thus makes type-II DSMs potential candidates for realizing novel topological superconductivity (TSC) different from other topological materials, which lacks investigations.

Higher-order topology, characterized by the presence of localized states (or charges) at $(d-m)$-dimensional ($m\geqslant 2$) boundaries in a $d$-dimensional material, has expanded the classification of topological phases of matter~\cite{HOTI,d-2,HOTI,CS1,CS2,CS3,CS4,2013HOTI,NW1,WN2}. The second-order topology in two dimension (2D) exhibits the quadruple moment and filling anomaly in a squared geometry~\cite{d-2,quadrupole,BBHPRB,Ben,WC1,WC2,WC3,WC4,quadrupolar_semimetal,GuoZhaopeng,Shao}. Furthermore, higher-order TSC~\cite{secondorder}, which is the combination of higher-order topology and superconductivity, is an intriguing field of research~\cite{Po_1,Po_2,theory_1,theory_2,theory_3,theory_4,theory_5,theory_6,theory_7,theory_8,theory_9,theory_10,theory_11,Dirac_SC,HOtypeI,PhysRevResearch.2.043155,Li_2022,PhysRevB.105.195149}. Some realistic models have been proposed in intrinsic materials~\cite{Das_Sarma,TRI_TSC}, through the proximity effect~\cite{MKP}, and with magnetic configurations~\cite{BHZ_TSC}. However, the higher-order TSC in type-II DSMs remains unexplored.

\begin{figure}[!t]
	\centering
	\includegraphics[trim=0 0 0 0, scale=1.25]{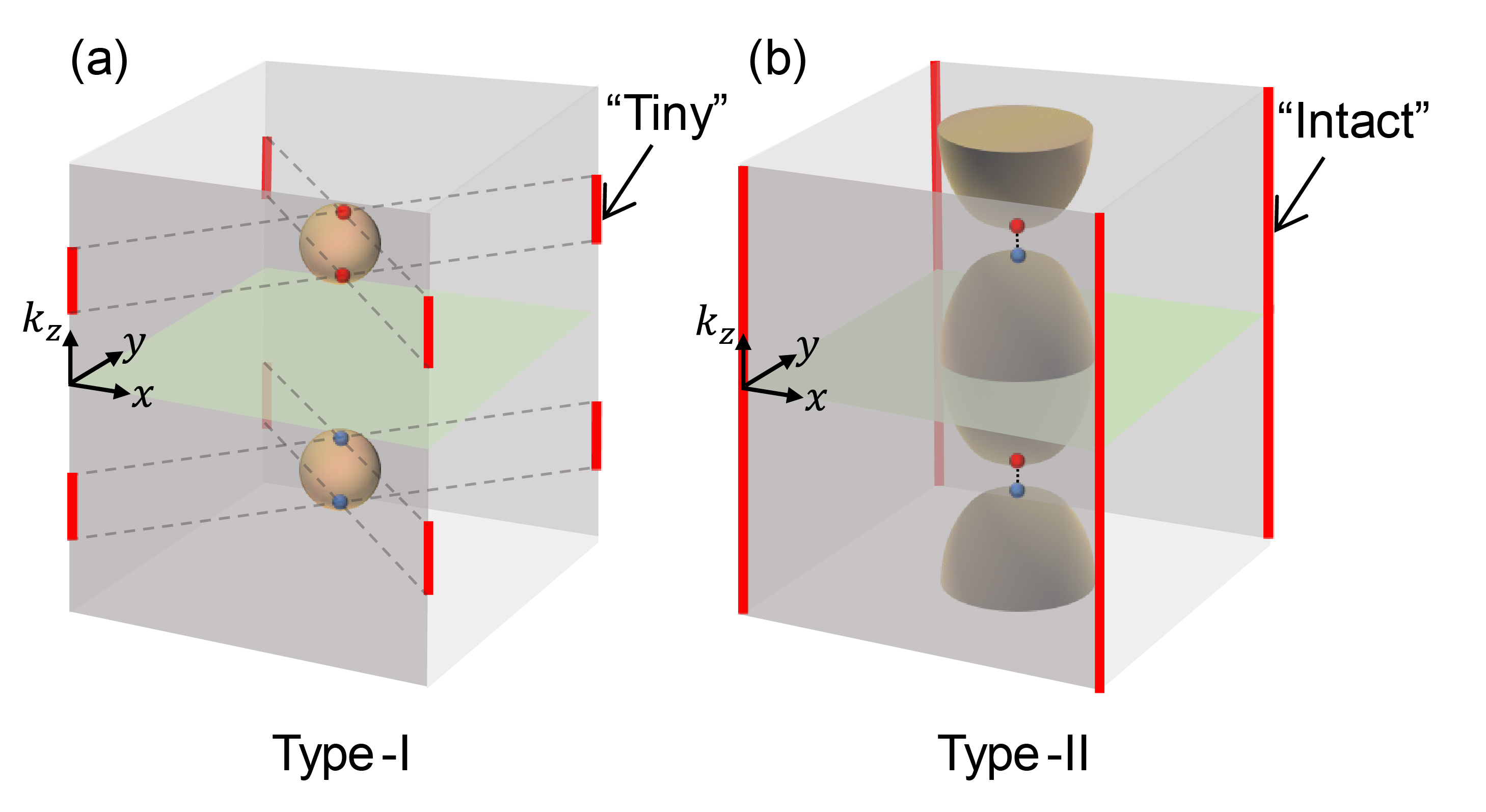}
	\caption{(a-b) Schematic plot of hinge Majorana flat bands (red lines) in superconducting type-$\I$ (a) and type-$\II$ (b) DSMs with dominant inter-orbital interactions. Fermi surfaces of the normal state are schematically plotted as yellow pockets. Superconducting Dirac points (red or blue dots) remain for the type-I, while they annihilate in pairs (depicted by dashed lines) and a full gap opens, resulting in a novel TSC state for the type-$\II$. The hinge Majorana states are indicated by red solid lines.}
	\label{fig1}
\end{figure}

In this letter, we reveal a novel topological superconducting state in a type-II DSM, which has a unique Majorana flat band on the hinge [Fig.~\ref{fig1}(b)]. By examining the interplay between intra- and inter-orbital interactions, we unveil the superconducting phase diagram for the type-II DSM. We find that when the inter-orbital attraction prevails, an unconventional odd-parity superconducting pairing is favored, resulting in a novel TSC state. It possesses both first- and second-order topology. Correspondingly, numerical simulations show the appearance of surface Majorana cones and hinge Majorana flat bands. Furthermore, we investigate the stability of the hinge Majorana modes against symmetry-breaking perturbations and the superconducting nodal structures.

\paragraph*{Model Hamiltonian.---} We consider a superconducting pairing in a type-II DSM of a tetragonal lattice with $D_{4h}$ symmetry, and the corresponding two-orbital Bogoliubov-de Gennes (BdG)  Hamiltonian reads:  
	\begin{equation}\label{Hk}\begin{aligned}
	H_{BdG}(\bk)& = \left(\begin{array}{cc}
		H_0(\bk)-\mu & D  \\
		D^\dagger & \mu-H_0^*(-\bk)   \\
	\end{array}\right),\\
	H_0(\bk)= 
& \{t_{\I}\cos{k_z}+t_{a}\left( \cos{k_x}+\cos{k_y} \right)-m\}\sigma_zs_0 \\
\ &+ t_{\II}\cos{k_z}\sigma_0s_0 + \eta\sin{k_x}\sigma_xs_z - \eta\sin{k_y}\sigma_ys_0\\
\ &+ \alpha\sin{k_z} \left( \cos{k_y}-\cos{k_x} \right)\sigma_xs_x\\
\ &+ \beta\sin{k_z}\sin{k_x}\sin{k_y}\sigma_xs_y.
	\end{aligned}\end{equation}
	\noindent The $4\cross 4$ Hamiltonian $H_0$ of the normal state describes a typical DSM on the basis of total angular momenta $J=\{3/2,\ 1/2\}$ with $C_{4z}$ eigenvalues $\{\mathrm{e}^{\pm \mathrm{i}\frac{3}{4}\pi},\mathrm{e}^{\pm \mathrm{i}\frac{1}{4}\pi}\}$ at $\Gamma$~\cite{Wang_Na3Bi,Wang_Cd3As2}, featuring a pair of Dirac points on the $k_z$ axis. The charge-conjugate operator is $\mathcal{C}=\tau_x\mathcal{K}$. $\bm{\tau}$, $\bm{\sigma}$ and $\bm{s}$ are Pauli matrices that act on particle-hole, orbital and spin subspaces, respectively. $m$ is the onsite energy difference between the two orbitals. $t_\I$, $t_a$, $t_\II$ are hopping amplitudes, and $\eta$, $\alpha$, $\beta$ are parameters related to spin-orbit coupling. The system has time reversal ($\mathcal{T}$), inversion ($\mathcal{P}$), fourfold rotation ($C_{4z}$) and mirror ($\mathcal{M}_x$) symmetries.


If $\abs{m-2t_a}<\abs{t_\I}$, this model hosts a pair of Dirac points, located at $\bk_D=(0,0,\pm k_0)$, with $k_0>0$ defined by $ k_0=\text{acos}[(m-2t_a)/t_\I]$. The presence of a finite $t_{\II}$ introduces tilting in the dispersion around these Dirac points. If $\vert t_{\II}\vert > \vert t_{\I}\vert$, Dirac cones are significantly tilted and become type-$\II$. Otherwise, they are of type-I. In the $k_z=0,\pi$ planes, an odd (even) number of $\mathcal{PT}$-pairs of FS sheets appear in type-$\II$ (type-$\I$) DSMs (Fig.~\ref{fig1}). This distinction in FS geometry is crucial to the hybrid TSC explored in type-$\II$ DSMs. 

\begin{table}[b]
\begin{center}
	\begin{tabular}{|c|c|c|c|c|c|c|}\hline
        Pairing channel & Matrix & Irrep. & $\ \mathcal{P}\ $ & $C_{4z}$ & $\mathcal{M}_z$ & $\mathcal{M}_x$ \\ \hline
		  $c_{1\uparrow}c_{1\downarrow}+c_{2\uparrow}c_{2\downarrow}$ & $D_{1a}=\Delta_{1a}\sigma_0s_y$ & \multirow{2}{*}{$A_{1g}$} & $+$ & $+$ & $+$ & $+$ \\
		  $c_{1\uparrow}c_{1\downarrow}-c_{2\uparrow}c_{2\downarrow}$ & $D_{1b}=\Delta_{1b}\sigma_zs_y$ &  & $+$ & $+$ & $+$ & $+$ \\
        \hline
		  $c_{1\uparrow}c_{2\uparrow}+c_{1\downarrow}c_{2\downarrow}$ & $D_{2}=\Delta_{2}\sigma_ys_0$ & $B_{1u}$ & $-$ & $-$ & $-$ & $-$ \\
        \hline
        $c_{1\uparrow}c_{2\uparrow}-c_{1\downarrow}c_{2\downarrow}$ & $D_{3}=\Delta_3\sigma_ys_z$ & $B_{2u}$ & $-$ & $-$ & $-$ & $+$ \\
        \hline
		  $c_{1\uparrow}c_{2\downarrow}-c_{1\downarrow}c_{2\uparrow}$ & $D_{4a}=\Delta_{4a}\sigma_xs_y$ & \multirow{2}{*}{$E_{u}$} & $-$ & \multirow{2}{*}{\diagbox[dir=SW]{\ }{\ }} & $+$ & $+$ \\
        $c_{1\uparrow}c_{2\downarrow}+c_{1\downarrow}c_{2\uparrow}$ & $D_{4b}=\Delta_{4b}\sigma_ys_x$ &  & $-$ &  & $+$ & $-$ \\
        \hline
    \end{tabular}
    \caption{Classification of all $k$-independent pairing potentials and their symmetry properties in the two-orbital $U$-$V$ model according to the representations of $D_{4h}$ point group.}\label{irreps}
\end{center}
\end{table}

\begin{figure}[!t]
	\centering
	\includegraphics[trim=10 0 0 0, scale=1.25]{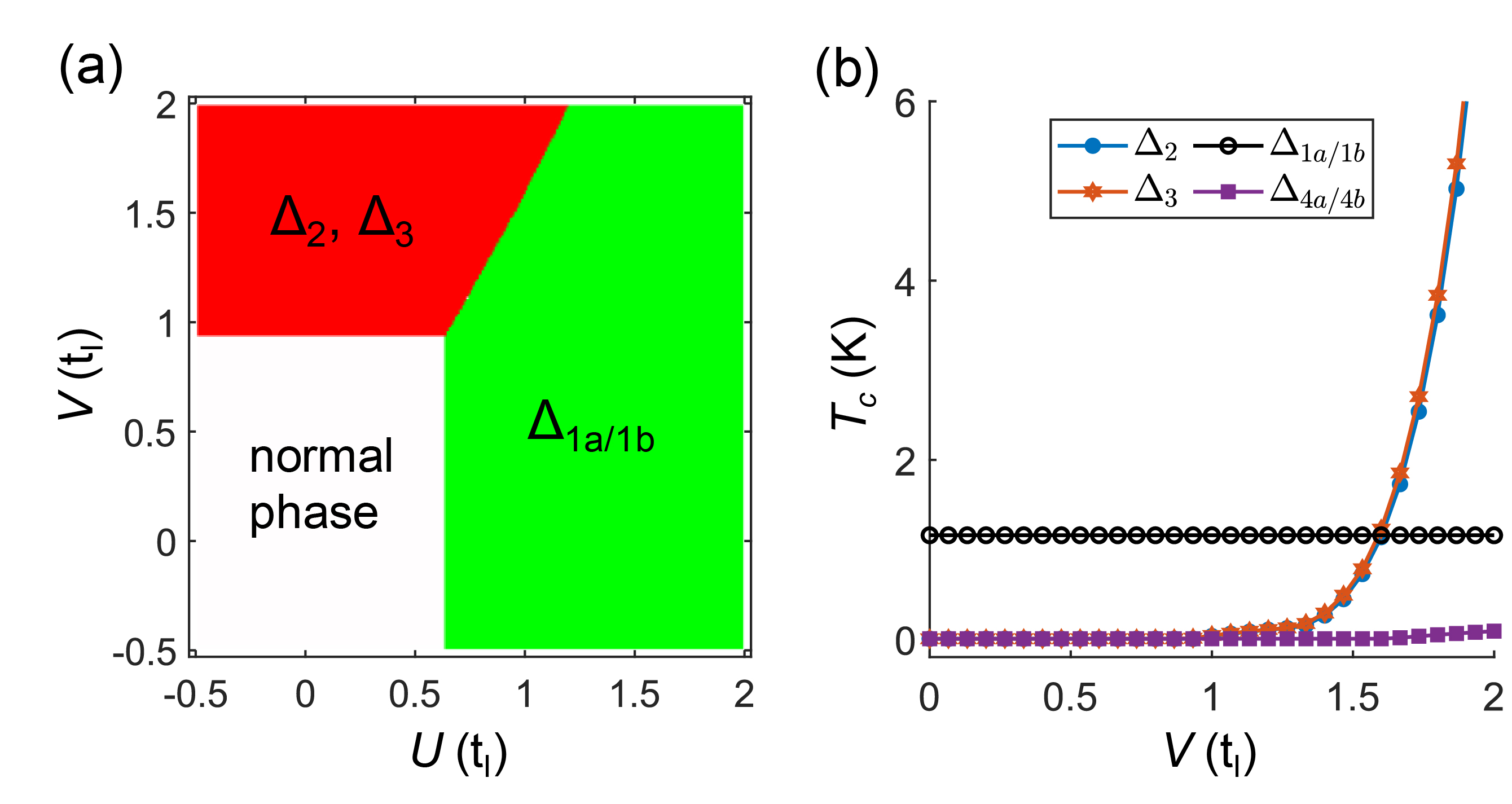}
	\caption{(a)~Phase diagram of the superconducting pairing channels. In the green region, the conventional $A_{1g}$ pairing, $\Delta_1$, is favored when the intra-orbital interaction $U$ dominates. In the red region, type-II DSMs favor the topological superconducting $B_{1u,2u}$ pairing channels, $\Delta_{2,3}$ when the inter-orbital interaction $V$ is sufficiently dominant. If the superconducting transition temperature falls below the threshold of $5\times10^{-3}$, the superconductivity is considered absent, and this state is identified as the normal phase (white region). (b)~Critical temperatures of various superconducting pairings as a function of the inter-orbital interaction $V$ with a fixed intra-orbital interaction $U=t_\I$. The critical temperature of the $E_u$ pairing remains consistently low and  the $B_{1u/2u}$ pairing is favored for $V>1.6 t_\I$. The adopted parameters are $m=4$, $t_\II=1.5$, $t_a=2$, $\eta=1.8$, $\alpha=1.2$, $\beta=2$, $\mu=0$, in the unit of $t_\I$. }
	\label{fig2}
\end{figure}

\paragraph*{Superconducting pairing channels.---}We study all possible $k$-independent pairing potentials and analyze their stability in the type-II DSM model~\cite{TSC_4,dopedDSM,TSC_2}. We now consider the following short-range density-density interaction in our model,
\begin{equation}
    H_{int}(x)=-U[n_1^2(x)+n_2^2(x)]-2Vn_1(x)n_2(x),
\end{equation}
where $U$ and $V$ describe intra- and inter-orbital interactions, respectively. There are six pairing terms satisfying the Fermionic anticommutation relation ($D=-D^\mathrm{T}$). Their irreducible representations and symmetry properties are listed in TABLE~\ref{irreps}. Among them, the $A_{1g}$ pairing channel is of even-parity and intra-orbital, while other pairing channels ($B_{1u}$, $B_{2u}$ and $E_u$) are of odd-parity and inter-orbital.\\
\indent We solve the linearized gap equation in the mean-field approximation [Section A of the Supplementary Material (SM)] to obtain critical temperatures for the pairing channels. The $U$-$V$ phase diagram is presented in Fig.~\ref{fig2}(a). The conventional $A_{1g}$ pairing, $\Delta_{1a/1b}$, is favored when the intra-orbital interaction $U$ is dominant (green area). However, with a dominant inter-orbital interaction $V$ (red area), the superconducting $B_{1u/2u}$ pairing channels $\Delta_2/\Delta_3$ are favored. These pairing states can also be promoted by electron-phonon coupling together with weak electronic correlations~\cite{TSC_6}. 



\indent In Fig.~\ref{fig2}(b), we plot the critical temperatures $T_c$ of different superconducting pairing channels as a function of the inter-orbital interaction $V$ ($U=t_\I$). As is shown, the $T_c$ of the $E_u$ pairing channel $\Delta_{4a,4b}$ is always the smallest among all pairing channels, so that it is suppressed and absent in the phase diagram. For small $V$, the superconductivity is dominant by conventional $A_{1g}$ pairing channel, $\Delta_{1a/1b}$. As $V$ increases, the $T_c$ of $\Delta_2$ and $\Delta_3$ pairings rise sharply for $V>1.3 t_\I$, surpassing the $A_{1g}$ channel at $V=1.6 t_\I$. These pairings become dominant at a larger $V$, and their close transition temperatures suggest the possibility of their coexistence. The phase difference $\Delta\phi$ between them is determined by the Josephson couplings in the free energy. Owing to their distinct symmetries, the first-order Josephson coupling vanishes and the second-order Josephson coupling dictates a phase difference of $\Delta\phi=\pm \pi/2$ to minimize the free energy. Hereafter, we consider the following pairing function,
\begin{equation}\label{paring}\begin{aligned}
	D=i\Delta_2\sigma_ys_0+\Delta_3\sigma_ys_z,
\end{aligned}\end{equation}
which allows us to discuss the TSC of both $B_{1u}$ and $B_{2u}$ channels in a unified framework.


\begin{figure}[!t]
	\centering
	\includegraphics[trim=5 0 0 0,scale=1.25]{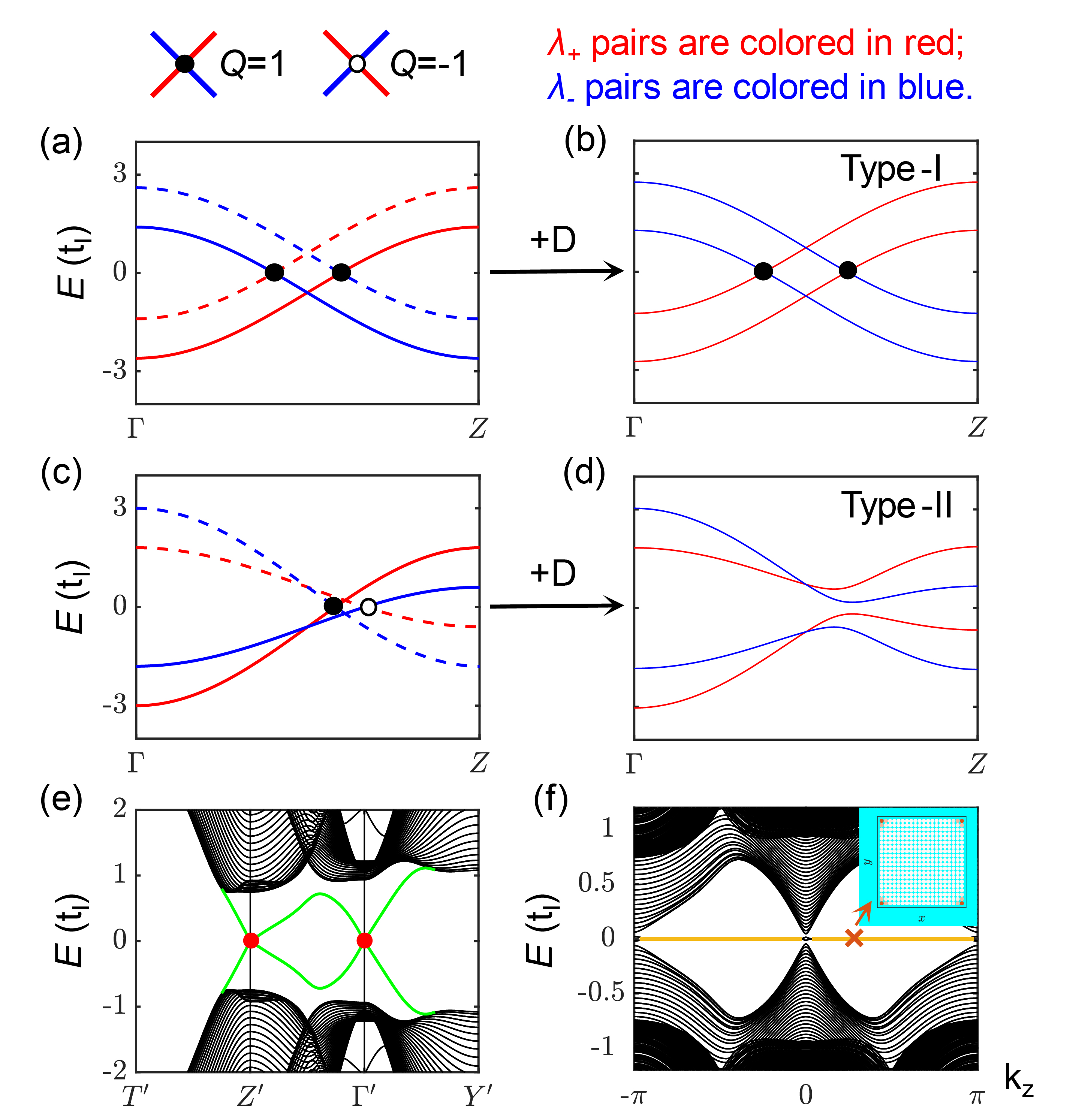} 
	\caption{ The bulk, edge and hinge dispersions of the BdG Hamiltonians of DSMs. (a,b) Bulk BdG dispersions of the type-I DSM without (a) and with (b) finite pairing potential $D$ in Eq.~(\ref{paring}). The BdG Dirac points remain.
    (c,d) Bulk BdG dispersions of the type-II DSM without (c) and with (d) pairing $D$. The BdG Dirac points annihilate when pairing potential is turned on, resulting in a novel full-gap TSC state.
    Dashed lines in (a,c) implies the hole part of the normal states.
    The red and blue lines indicate the different pairs of bands with the BdG $\tilde{C}_{4z}$ eigenvalues ($\lambda_\xi=\xi \mathrm{e}^{\pm \mathrm{i}\frac{\pi}{4}}$, $\xi=\pm $), respectively.
    The surface (e) and hinge (f) dispersions of the novel TSC state. In (e), the surface Majorana states at $\Gamma'$ and $Z'$ are dipicted by red dots. Two helical Majorana cones are obtained on the (100) surface. In (f), surprisingly, a Majorana flat band (highlighted in orange) spans the entire hinge Brillouin zone and connects the projections of the surface helical Majorana cones. The inset shows the real-space wave function distribution of the hinge modes for the $k_z = \pi/4$ plane with four corner MZMs localized around the four corners. In both (e) and (f), the lattice Hamiltonian in the Eq.~(\ref{Hk}) is adopted with pairing potenrial Eq.~(\ref{paring}), $(\Delta_2, \Delta_3) = (1.2, 0.2)$. }
	\label{fig3}
\end{figure}

\paragraph*{First-order TSC and helical Majorana surface states for $k_z=0,\pi$ planes.---} 
Due to the odd-parity pairing nature, $\mathcal{P}D\mathcal{P}^\mathrm{T}$=$-D$, the topological invariant is determined by the number of FS pairs (Kramers degeneracy) of the normal state~\cite{TSC_1,TSC_2,TSC_3}. Here in our model, for the $\mathcal{T}$-invariant $k_z=0,\pi$ plane, in the type-II DSM region, an odd number of pairs of FS sheets that enclose $\mathcal{T}$-invariant momenta results in a nontrivial $\mathbb{Z}_2$ topology of the DIII class~\cite{AZ_class}. This makes the $k_z=0,\pi$ plane of a type-II DSM becomes a first-order $\mathcal{T}$-invariant 2D topological superconductor. Consequently, with open boundary conditions, surface Majorana states at the $\Gamma'$ and $Z'$ emerge on the (100) surface, as illustrated in Fig.~\ref{fig3}(e).

\paragraph*{Second-order TSC and the hinge Majorana flat bands.---}For those 2D $k_z$-planes other than $0$ or $\pi$,  time-reversal symmetry is broken and edge states become gapped, trivializing the first-order topology. However, we find that they exhibit the second-order topology, characterized by corner Majorana zero modes (MZMs). These corner MZMs of different $k_z$ planes assemble to form an intact Majorana flat band at each $z$-directed hinge of a squared rod geometry. The hinge Majorana flat band penetrates through the entire hinge Brillouin zone and links the projections of the surface helical Majorana cones at $k_z=0,\pi$ points, as shown in Fig.~\ref{fig3}(f). The second-order topology is characterized by using the nested Wilson loop method, as detailed in Section D of the SM.

\begin{figure}[!tb]
	\centering
	\includegraphics[trim=8 0 0 0, scale=1.2]{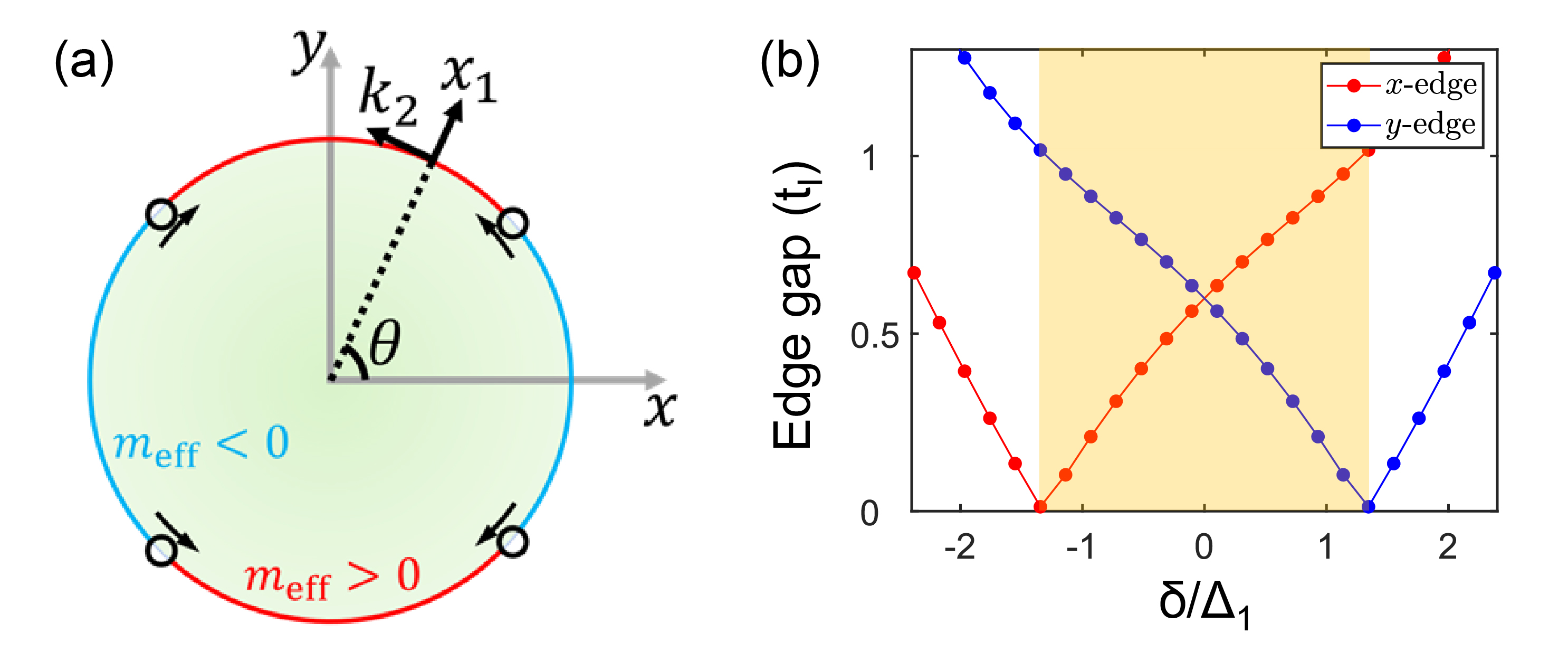}
	\caption{(a) Schematic plot of a circular geometry with an arbitrary edge labeled by angle $\theta$ (purple tangent line). In the red (blue) region, the effective edge mass $m_{\mathrm{eff}}(\theta)$ is positive (negative). At the black hollow dots, $m_{\mathrm{eff}}(\theta)$ vanishes and undergoes a sign change. A $C_{4z}$-symmetry-breaking perturbation
moves these points toward each other, altering the blue region's size as depicted by black arrows, until they annihilate at the edges with	$\theta=0,\pi$ or $\theta=\frac{\pi}{2},\frac{3\pi}{2}$. (b) Evolution of the $x$- and $y$-edge gaps of the $k_z=\pi/4$ plane with the $C_{4z}$-symmetry-breaking perturbation $\delta$. The yellow region indicates the topologically nontrivial phase where $x$- and $y$-edge gaps have opposite signs, leading to corner Majorana zero modes.}
	\label{fig4}
\end{figure}

\paragraph*{Stability of Majorana hinge modes.---}
The novel TSC state can be understood from the low-energy analysis. 
Starting from the 8$\times$8 $H_{BdG}(\bk)$ in Eq.~(\ref{Hk}), we extract a 4$\times$4 low-energy continuum Hamiltonian for each $k_z$ plane as follows (see Section B of the SM for details):

\begin{equation}\label{massful_pxpy}\begin{aligned}
	 H_{k_z\text{-plane}}(\bkp) &= \left(\begin{array}{cc}
		-M(\bkp,k_z) \mathbb{I}_{2\times 2} & \mathbb{A}_{2\times 2} \\
		\mathbb{A}^{\dagger}_{2\times 2}& M(\bkp,k_z)\mathbb{I}_{2\times 2} \\
		\end{array}\right),\\
  \rm { with~}\mathbb{A}_{2\times 2}&= \left(\begin{array}{cc}
		 - \Delta_-k_- & S(\bkp,k_z) \\
		S(\bkp,k_z) &  \Delta_+k_+ \\
		\end{array}\right).
\end{aligned}\end{equation}
Here $M(\bkp,k_z)=-(t_\II- t_{\I})\cos k_z + \frac{1}{2}(\frac{\eta^2}{t_{\I}\cos k_z}-t_{a})\bkpsq$, $\Delta_\pm=\eta\frac{\Delta_{2}\pm i\Delta_3}{t_{\I}\cos k_z}$, $k_\pm=k_x\pm ik_y$ and $\bkp=(k_x,k_y)$. Here we take $m = 2 t_a$ to avoid complexity in the derivation. This continuum Hamiltonian is a spinful $p_x$+$ip_y$-like model~\cite{pxpy}, decorated by an additional $\mathcal{T}$-breaking term $S(\bkp,k_z)=-\sin k_z\left[ \gamma_2(k_x^2-k_y^2)+2\gamma_3k_xk_y \right]$, with $\gamma_2=\frac{\alpha\Delta_2}{(t_\I+ t_{\II})\cos k_z}$ and $\gamma_3=\frac{\beta\Delta_3}{(t_\I+ t_{\II})\cos k_z}$.
For $\mathcal{T}$-invariant $k_z=0,\pi$ plane, $S(\bkp,k_z)=0$ and Eq.~(\ref{massful_pxpy}) gives rise to the $\mathcal{T}$-protected first-order TSC. For the $k_z\neq 0,\pi$ plane, $S(\bkp,k_z)\neq0$ and the edge states of the 2D model are gapped. The gaps strongly depend on the direction of the edge and thus are anisotropic. To visualize this, we first notice that an edge termination can be labeled by the angle $\theta$ in a dish geometry in Fig.~\ref{fig4}(a). By imposing open boundary conditions on Eq.~(\ref{massful_pxpy}) under the local coordinate, we obtain the effective edge Hamiltonian with the angle $\theta$ (see Section C of SM)~\cite{33_from_Das,54_from_Das}:
\begin{equation}\label{edge}\begin{aligned}
	&H_{\theta}^{\mathrm{edge}}(k_2) = \vert\Delta\vert k_2\bar{\sigma}_z + m_{\mathrm{eff}}(\theta)\bar{\sigma}_y.
\end{aligned}\end{equation}
\noindent $\bar{\bsigma}$ are Pauli matrices acting on the two low-energy edge states. $k_2$ is the momentum along the edge and $\vert \Delta \vert = \sqrt{\Delta_+\Delta_-}$. The edge gap term is $m_{\mathrm{eff}}(\theta) \propto \alpha\Delta_2\cos{2\theta}+\beta\Delta_3\sin{2\theta}$. Notably, the angular anisotropy of $m_{\mathrm{eff}}(\theta)$ enables the existence of four mass-sign flipping domain walls of edges, where the corner MZMs exist.

Specifically, the fourfold rotation operator on the edge is $C_{4z}^{\mathrm{edge}}(\theta)=-i\bar{\sigma}_z$ and the mass term in Eq.~(\ref{edge}) satisfies $C_{4z}^{\mathrm{edge}}(\theta)m_{\mathrm{eff}}(\theta)\bar{\sigma}_y	C_{4z}^{\mathrm{edge} \dagger}(\theta)=m_{\text{eff}}(\theta+\pi/2)\bar{\sigma}_y$. Therefore, one must have $m_{\text{eff}}(\theta+\pi/2)=-m_{\text{eff}}(\theta)$, which means that the fourfold rotation enforces the existence of mass domains at the corresponding corners and leading to the emergence of corner MZMs.

Unconventional superconductivity is always accompanied with other symmetry-breaking orders, such as nematicity, spin or charge density wave~\cite{IrTe2,IrTe2_2}. We study the stability of the corner MZMs under the $C_{4z}$-breaking perturbation by phenomenologically introducing a symmetry-breaking term $\delta\sin{k_z}\sigma_xs_x$ to $H_0(\bk)$ in Eq.~(\ref{Hk}). Considering $\Delta_3=0$, this results in an additional mass term $m_\delta\propto\delta\Delta_2\sin{k_z}$ to the edge dispersion of Eq.~(\ref{edge}). The $C_{4z}$-breaking $\delta$, preserving the mirror symmetries $\mathcal{M}_x$ and $\mathcal{M}_y$, allow two pairs of the corner MZMs [green dots in Fig.~\ref{fig4}(a)] to move towards and finally annihilate with each other at $x=0$ or $y=0$. The corner MZMs are still topologically protected at the boundaries between different mass regions. The additional $C_{4z}$ symmetry does nothing but pins the corner MZMs on $x=\pm y$ lines. Fig.~\ref{fig4}(b) shows energy gaps of $x$- and $y$-edges for $k_z=\pi/4$ plane as a function of $\delta$ (in units of $\Delta_2$). The yellow shaded region shows the higher-order nontrivial region with the existence of corner MZMs. The mass term $m_\mathrm{eff}(\theta)\propto\alpha\Delta_2\cos{2\theta}+\beta\Delta_3\sin{2\theta}$ indicates that the nontrivial region linearly depends on the spin-orbit coupling strength ($\alpha$ and $\beta$). Consequently, enhancing the spin-orbit coupling can broaden the stability range of hinge Majorana modes.



\paragraph*{Superconducting nodal structure.---}In the bulk BdG dispersion, there is parameter region that superconducting Dirac points (SC DPs) appear. The SC DPs can be characterized by the eigenvalues of the generalized fourfold rotation $\tilde{C}_{4z}$, acting on the full BdG space. For a gap function of Eq.~(\ref{paring}) with $C_{4z}DC_{4z}^\mathrm{T}=- D$, the fourfold rotation is represented by $\tilde{C}_{4z}=\text{diag}\{C_{4z},-C_{4z}^*\}$. On the $\tilde{C}_{4z}$-invariant line $\Gamma Z=(0,0,k_z)$, the BdG Hamiltonian commutes with $\tilde{C}_{4z}$. Due to the presence of the $\mathcal{P}\mathcal{T}$ symmetry, all the bands are doubly degenerate and have a pair of $\tilde{C}_{4z}$ eigenvalues ($\lambda_+=\mathrm{e}^{\pm \mathrm{i} \frac{\pi}{4}}$ or $\lambda_-=-\mathrm{e}^{\pm \mathrm{i} \frac{\pi}{4}}$). The particle-hole symmetry related bands are of the different pairs, resulting in a $\tilde{C}_{4z}$-symmetry protected SC DP at the Fermi level.  We defined  $Q=+1$ (resp. $Q=-1$) for the SC DP formed by the BdG bands of the positively-sloped  $\lambda_+$ (resp. $\lambda_-$) pair and the negatively-sloped $\lambda_-$ (resp. $\lambda_+$) pair. One can conclude that only the opposite-$Q$ SC DPs can annihilate.

As shown in Fig.~\ref{fig3}(a), for a type-I DSM, along the $\Gamma$-$Z$ line in the BdG band dispersion in the weak coupling limit ($D=0$), there are two SC DPs with identical $Q$ numbers~\cite{Sato_TCSC}. They are stable after turning on the pairing potential [Fig.~\ref{fig3}(b)]. Tiny hinge Majorana states are found between the projection of the bulk SC DPs in Fig.~\ref{fig1}(a)~\cite{HOtypeI}. However, in Fig.~\ref{fig3}(c) in weak coupling limit, the $Q$ numbers of two adjacent SC DPs are opposite in a type-II DSM. Upon turning on the superconducting pairing [Fig.~\ref{fig3}(d)], the two SC DPs can meet and annihilate with each other, resulting in a full BdG gap, which allows the existence of the novel TSC state with intact hinge Majorana flat bands spanning the entire hinge Brillouin zones.

\paragraph*{Discussion and conclusions.---} In summary, we propose a novel TSC state in a type-II DSM, induced by the dominant inter-orbital attraction. This novel state has an unconventional odd-parity superconducting pairing, giving rise to the novel TSC, which is characterized by the coexistence of surface Majorana cones and hinge Majorana flat bands. Moreover, studies in Refs.~\cite{s_F_p,s_p,s_p_Experiment}  indicate that in a superconductor junction, the $T_c$ of an odd-parity superconductor can be enhanced by up to 1.5 times. Our proposal would intrigue researchers to search for the hinge Majorana flat bands of the novel TSC state in experiments.


\ \\
\noindent \textbf{Acknowledgments}
This work was supported by the Strategic Priority Research Program of Chinese Academy of Sciences (Grant No. XDB33000000), the National Natural Science Foundation of China (Grants No. 12188101, No. 52188101 and No. 11974395), and the Center for Materials Genome.

\bibliography{Refs}

\ \\

\clearpage

\begin{widetext}

        \beginsupplement{}
        \setcounter{section}{0}
		 \setcounter{equation}{0}
        \renewcommand{\thesubsection}{\Alph{subsection}}
        \renewcommand{\thesubsubsection}{\alph{subsubsection}}
		 \renewcommand{\theequation}{S\arabic{equation}}
\section*{SUPPLEMENTARY MATERIAL}

\subsection{Linearized gap equation of the possible pairing channels}
In this section, we are going to demonstrate the linearized gap equation method to obtain the critical temperatures of the possible pairing channels. First recall the main Hamiltonian of type-II Dirac semimetal (DSM) in the main text and define some abbreviations, 
	\begin{equation}\label{HinS}\begin{aligned}
	H_0(\bk)&= t_{\II}\cos{k_z}\sigma_0s_0 + \{t_{\I}\cos{k_z}+t_{a}\left( \cos{k_x}+\cos{k_y} \right)-m\}\sigma_zs_0 
+ \eta\sin{k_x}\sigma_xs_z - \eta\sin{k_y}\sigma_ys_0\\
&\quad+ \alpha\sin{k_z} \left( \cos{k_y}-\cos{k_x} \right)\sigma_xs_x
+ \beta\sin{k_z}\sin{k_x}\sin{k_y}\sigma_xs_y\\
& \equiv -\mu_\II(\bk)\Gamma_0+a(\bk)\Gamma_1+b(\bk)\Gamma_2+c(\bk)\Gamma_3+d(\bk)\Gamma_4+e(\bk)\Gamma_5
	\end{aligned}\end{equation}
$\Gamma_0$ is $4\times 4$ identity and $\Gamma_1=\sigma_z\otimes s_0$, $\Gamma_2=\sigma_x\otimes s_z$, $\Gamma_3=\sigma_y\otimes s_0$, $\Gamma_4=\sigma_x\otimes s_x$, $\Gamma_5=\sigma_x\otimes s_y$, $\Gamma_{ij}=\left[\Gamma_i,\Gamma_j\right]/2i$. $\bsigma$ and $s$ are Pauli matrices describing the orbital and spin degrees of freedom, respectively. The system has time reversal ($\mathcal{T}$), inversion ($\mathcal{P}$), mirror ($\mathcal{M}_z$) and fourfold rotation ($C_{4z}$) symmetries as below,
	\begin{equation}\begin{aligned}
	\mathcal{T}=-is_y\mathcal{K},\quad \mathcal{P}=-\sigma_z,\quad
	\mathcal{M}_z=-is_z,\quad C_{4z}=\exp\left[i(\pi/4)(2+\sigma_z)\otimes s_z\right].
	\end{aligned}\end{equation}
The bands energy dispersion is $E_\pm(\bk)=\pm E_0(\bk)-\mu_\II(\bk)$ with $E_0=\sqrt{a^2+b^2+c^2+d^2+e^2}$. It is clear that the parameter associate with the identical term, $\mu_\II(\bk)=-t_{\II}\cos{k_z}$, can be regarded as a $\bk$-dependent chemical potential.

Considering the following short-range density-density interaction,
\begin{equation}
    H_{int}(x)=-U[n_1^2(x)+n_2^2(x)]-2Vn_1(x)n_2(x)
\end{equation}
where $U$ and $V$ are intra- and inter-orbital interaction strength. $n_\sigma=n_{\sigma\uparrow}+n_{\sigma\downarrow}$ and $\sigma=1,2$ denotes orbitals with total orbital angular momentum 1 and 0, respectively. For our two-orbital model, in the mean-field level, there are six possible pairing potentials that satisfy the Fermi-Dirac statistics. Expressed in the combinations of two Pauli matrices, they are $D_{1a}=\Delta_{1a}\sigma_0s_y$, $D_{1b}=\Delta_{1b}\sigma_zs_y$, $D_{2}=\Delta_{2}\sigma_ys_0$, $D_{3}=\Delta_{3}\sigma_ys_z$, $D_{4a}=\Delta_{4a}\sigma_xs_y$, $D_{4b}=\Delta_{4b}\sigma_ys_x$. Here, $\Delta_{1a,1b}$, $\Delta_2$, $\Delta_3$ and $\Delta_{4a,4b}$ are of $A_{1g}$, $B_{1u}$, $B_{2u}$ and $E_u$ pairing channels.

Applying the linearized gap equation to possible pairing channels, we have for the $A_{1g}$ channel,
\begin{equation}\label{lge1}
    \left|\begin{array}{cc} U\chi_{1a}-1 & U\chi_{1ab} \\ U\chi_{1ab} & U\chi_{1b}-1 \end{array}\right|=0,
\end{equation}
and for other channels ($B_{1u}$, $B_{2u}$, $E_u$)
\begin{equation}\label{lge2}
    V\chi_j = 1,
\end{equation}
with $j=2,3,4a,4b$. Here, $\chi_j$'s are finite temperature superconducting susceptibilities in different pairing channels. By defining the single-particle normal-state Green function $G_0(\bk,i\omega_n)=\sum_{m=\pm}P_m(k)/\{i\omega_n-[mE_0(k)-\mu_{\II}(k)]\}$, with the projection operator $P_{+(-)}$ onto the higher (lower) energy bands,
\begin{equation}
    P_\pm(\bk)=\frac{1}{2}(1\pm\frac{H_0(\bk)+\mu_\II(\bk)}{ E_0(\bk)}),
\end{equation}
the superconducting susceptibility is written as,
\begin{equation}\label{susceptibility}\begin{aligned}
    \chi_j &=\frac{1}{2N\beta_c}\sum_k\sum_{i\omega_n}\mathrm{Tr}\{M_\Delta G_0(\bk,i\omega_n)M_{\Delta'} G_0^T(-\bk,-i\omega_n)\}\\
		&=\frac{1}{N}\sum_{m=\pm}\sum_k F_j^m(\bk)\frac{\tanh{\frac{1}{2}\beta_c[mE_0(\bk)-\mu_\II(\bk)]}}{2[mE_0(\bk)-\mu_\II(\bk)]}.
\end{aligned}\end{equation}
Here, $M_\Delta$ is the matrix part of the pairing potential and $\beta_c=1/k_BT_c$. Note that type-$\II$ Dirac semimetals are different from type-$\I$ ones that they have both electron and hole pockets centering around the center and the boundary of the Brillouin zone, respectively contributed by higher and lower energy bands. This gives rise to the summation $m=\pm$ in Eq.~(\ref{susceptibility}). $F_j^\pm(\bk)$ are the form factors and is given by $F_{1a}^\pm(\bk)=1$, $-F_{1ab}^-(\bk) = F_{1ab}^+(\bk) = \frac{a(k)}{E_0(k)}$, $F_{1b}^\pm(\bk)=\frac{a^2(k)}{E_0^2(k)}$, $F_2^\pm(\bk)=\frac{b^2(k)+c^2(k)+d^2(k)}{E_0^2(k)}$, $F_3^\pm(\bk)=\frac{b^2(k)+c^2(k)+e^2(k)}{E_0^2(k)}$, $F_{4a}^\pm(\bk)=\frac{b^2(k)+d^2(k)+e^2(k)}{E_0^2(k)}$ and $F_{4b}^\pm(\bk)=\frac{c^2(k)+d^2(k)+e^2(k)}{E_0^2(k)}$. From the form factors, it is obvious that $\chi_{2,3}>\chi_{4a,4b}$ since $d^2(\bk)$ and $e^2(\bk)$ are $k^3$ terms, so that the $E_u$ channel will not dominant for any choices of U and V. Around Fermi surface, take $E_0(\bk)\approx \mu_\II(\bk)$, and in the U-V phase diagram, the phase boundary between $D_1$ and $D_{2,3}$ is approximately given by~\cite{dopedDSM},
\begin{equation}\begin{aligned}
    \frac{V}{U}=\int_{FS\pm}d\bk\frac{\mu^2_\II(\bk)+a^2(\bk)}{\mu_\II^2(\bk)-a^2(\bk)}.
\end{aligned}\end{equation}
For type-I DSM, the phase boundary is approximitely $V/U\approx 2.1$~\cite{dopedDSM}. For type-II Dirac semimetal, at the limit $t_\II\gg t_\I$ and thus $\mu^2_\II(k)\gg a^2(k)$, the $B_{1u}/B_{2u}$ region will be broaden to $V/U\approx 1$. The critical temperatures of different pairing channels can be solved numerically using Eq.~(\ref{lge1}, \ref{lge2}).

Additionally, From the form factors, we note that $F_{2,3}^\pm$ vanishes at $k_x=k_y=0$, which are the polar points of the Fermi surfaces. Therefore, these $\bk$ points do not participate in pairing and they will become gapless nodal points in BdG band dispersion if the pairing is weak. However, as we shall see, these nodal points have different monopole charges and thus will annihilate each other to give rise to a fullgap superconductor.

\subsection{Low-energy analysis}
In this section, we are going to extract a low-energy Hamiltonian around the Fermi surfaces to describe the hybrid-order topological superconductivity (\ie the coexistence of first- and second-order topological superconductivity). For simplicity, we would investigate each $k_z$-fixed plane individually. With fixed $k_z$, the continuous 2D-plane Hamiltonian is, 
	\begin{equation}\begin{aligned}\label{continuous}
	&H^{\mathrm{2D}}(\bkp) = \left(\begin{array}{cc}
		H_0^{\mathrm{2D}}(\bkp)-\mu & D  \\
		D^\dagger & \mu-H_0^{\mathrm{2D}*}(-\bkp)   \\
	\end{array}\right),\\
	 &H_0^{\mathrm{2D}}(\bkp) = \tilde{t}_{\II}\Gamma_0 + \tilde{t}_{\I}\Gamma_1 - \frac{t_{a}}{2}(k_x^2+k_y^2)\Gamma_1 \\&\qquad\qquad+ \eta(k_x\Gamma_2-k_y\Gamma_3) + \frac{1}{2}\tilde{\alpha}(k_x^2-k_y^2)\Gamma_4+\tilde{\beta}k_xk_y\Gamma_5,\\
	&\tilde{t}_{\II,\I}=t_{\II,\I}\cos k_z,\ \tilde{\alpha}=\alpha\sin{k_z},\ \tilde{\beta}=\beta\sin{k_z},
	\end{aligned}\end{equation}
\noindent  where, $\bkp=(k_x,k_y)^T$. Here we take $m = 2 t_a$ to avoid complexity in the derivation. We now take $\tilde{t}_{\I,\II}>0$ without loss of generality. In a system with helical texture, the low energy downfolding using helical basis $\psi_k=(c_{\uparrow k}+e^{i\theta_k}c_{\downarrow k})/\sqrt{2}$ yields effective typical topological superconductor model~\cite{Fu_Kane,Alicea}. Here in our model, the Fermi surfaces of the normal state have helical orbit-momentum locking, as shown in Fig.~\ref{figS1}(a). When orbit-momentum locking is strong enough ($\eta^2>\tilde{t}_\I\vert t_a\vert$) and assuming weak pairing assumption ($k_z \simeq 0,\pm\pi$), a pair of Kramers degenerate FS sheets can be very well described by a low-energy dwonfolding by apply unitary transformation $U(\bkp)\equiv(\Psi_\bk,\sigma_x\Psi^*_{-\bk})$. Here, $\Psi_\bk$ are the four low-energy BdG bands near the Fermi surface and $\sigma_x\Psi^*_{-\bk}$ are high-energy BdG bands. However, such transformation in our model is more complicated, and it reads:
	\begin{equation}\begin{aligned}
	\Psi_\bk &= \left(\begin{array}{cccc} \varphi_\bk & -is_y\varphi^*_{-\bk} & \tau_x\varphi^*_{-\bk} & -i\tau_xs_y\varphi_{\bk} \end{array}\right),\\
	\varphi_\bk &= \left(\begin{array}{cccccccc} 
		A_\bk & 0 & B_\bk & 0 & 0 & 0 & 0 & 0 
	\end{array}\right)^T/\sqrt{\mathcal{N}_\bk}.
	\end{aligned}\end{equation}
\noindent Here, $B_\bk=2\tilde{t}_{\I}-t_{a}\bkpsq+\sqrt{(-2\tilde{t}_{\I}+t_{a}\bkpsq)^2+4\eta^2\bkpsq}$ and $A_\bk=-2\eta(k_x+ik_y)$, with the normalization factor $\mathcal{N}_\bk$. One can see that $A_\bk$ represents the helicity of the orbit-momentum locking. $\bm{\tau}$ is Pauli matrix describing the particle-hole degree of freedom. Note that $-is_y\varphi^*_{-\bk}$ is the time-reversal (TR) counterpart of $\varphi_\bk$ and $\tau_x\varphi^*_{-\bk}$ is the charge-conjugate counterpart of $\varphi_\bk$. Now, the effective Hamiltonian is
	\begin{equation}\begin{aligned}
	&H_{\mathrm{eff}}(\bkp)
	= U^\dagger H^{\mathrm{2D}}(\bkp) U
	= \left(\begin{array}{cc}
		H_l(\bkp) & H_s(\bkp) \\
		H^\dagger_s(\bkp) & H_h(\bkp)
	\end{array}\right).\\
	&H_l(\bkp) = \left(\begin{array}{cccc}
		-M + T\bkpsq & 0 & - \tilde{\Delta}_-k_- & 0 \\
		0 & -M + T\bkpsq & 0 &  \tilde{\Delta}_+k_+ \\
		- \tilde{\Delta}_+k_+ & 0 & M - T\bkpsq & 0 \\
		0 &  \tilde{\Delta}_-k_- & 0 & M - T\bkpsq \\
		\end{array}\right)
	\end{aligned}\end{equation}
\noindent Here, $H_l(\bkp)$ is the effective Hamiltonian in low-energy space, with $M=\tilde{t}_{\I}-\tilde{t}_{\II}$, $T=\left(t_{a}-\eta^2/\tilde{t}_{\I}\right)/2$, $\tilde{\Delta}_\pm=\eta(\Delta_{2}\pm i\Delta_3)/\tilde{t}_{\I}$ and $k_\pm=k_x\pm ik_y$. Low-energy space is spanned by $\tilde{\bm{\tau}}\otimes\tilde{\bm{s}}$, respectively describing pseudo-particle-hole and -spin degrees of freedom. The low-energy physics is clearly a TR-invariant $p_x$+$ip_y$-like model~\cite{pxpy}, originating from both orbital texture and inter-orbital pairing. By Fu-Kane criterion~\cite{Fu_Kane_criterion}, edge-localized states always exist for there is one pair of FS sheets in a type-II DSM. For the TR-invaiant plane ($k_z=0\ or \ \pi$), nontrivial $\mathbb{Z}_2$ phase ensures such localized states to be helical Majorana states.
\\
\indent For $k_z\neq 0\ or\ \pi$ planes, however, due to the lack of time reversal symmetry that square to $-1$, the SOC term will gap out the edge states. More specifically, one can perform a higher-order perturbation by introducing scattering with high-energy bands $H_h$ through Green's function $G(\bkp,\omega)=( \omega-H_{\mathrm{eff}})^{-1}$. The correction to $H_l$ is $H'_l(\bkp,\omega)=H_s\left(\omega-H_h\right)^{-1}H^\dagger_s$. Choose $\omega=0$ as an approximation, the corrected low-energy bands becomes,
	\begin{equation}\begin{aligned}\label{fullHeff}
	H_{k_z\text{-plane}}(\bkp) = H_l(\bkp)+H_l'(\bkp,0).
	\end{aligned}\end{equation}
\noindent $H'_l(\bkp,0)=-\left[ \gamma_1(k_x^2-k_y^2)+2\gamma_2k_xk_y \right]\Gamma_4$ is anti-diagonal, with $\gamma_1=\tilde{\alpha}\Delta_2/(\tilde{t}_{\I}+\tilde{t}_{\II})$ and $\gamma_2=\tilde{\beta}\Delta_3/(\tilde{t}_{\I}+\tilde{t}_{\II})$. One can see that the $\bk$-independent pairing $D$ enters into the low-energy bands mediating through SOC as an \textit{inherited pairing} $H'_l(\bkp,0)$, \ie an effective pairing inherits the functional distribution of the SOC term, and becomes distributed. Interestingly, $B_{1u}$ and $B_{2u}$ pairing channels enter the low-energy bands separately through $\alpha$ and $\beta$ terms. Additionally, it can be seen from Eq.~(\ref{fullHeff}) that the pairing potential vanishes on $\Gamma$-$Z$, and it will leave two nodes on the Fermi surface ungapped at weak pairing limit. However, as discussed in the main text, the nodes are of opposite monopole charges and will vanish eventually when increasing pairing strength. This is consistent with the form factors (Eq.~\ref{susceptibility}) obtained from the linearized gap equation.

\begin{figure}[t!]
	\centering
	\includegraphics[scale=1.25]{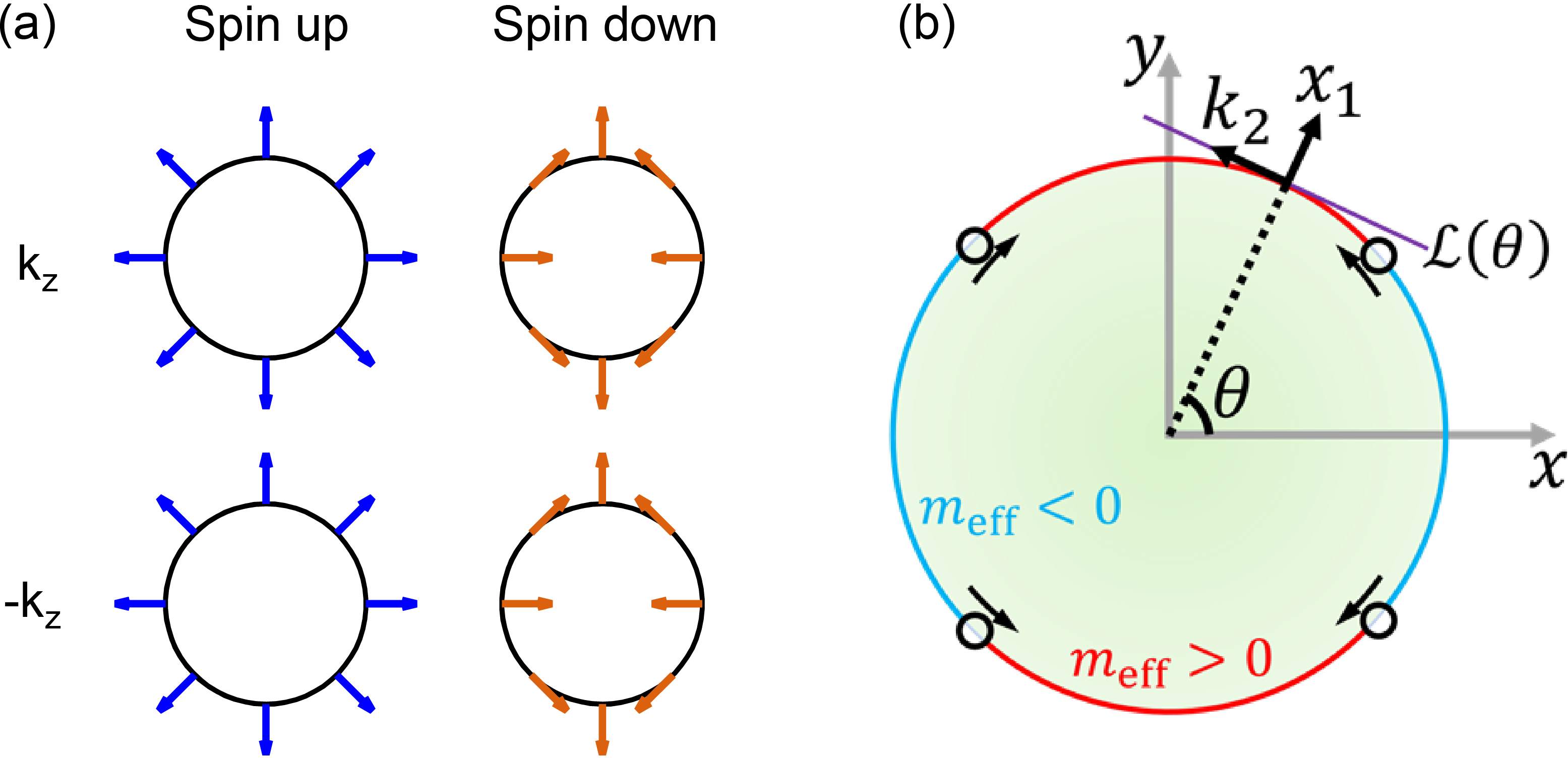}
	\caption{(a) Orbital texture on Fermi surface with fixed-$k_z$ planes for spin-up (left) and spin-down (right) sectors. Arrows on Fermi surfaces indicate the orbit-momentum texture ($\ev{\sigma_x},\ev{\sigma_y}$). (b) Schematic plot of a circular geometry with arbitrary edge $\mathcal{L}(\theta)$ (purple tangent line). In the red (blue) region, the effective edge mass $m_{\mathrm{eff}}(\theta)$ is positive (negative). The black hollow dots denote the corner Majorana zero modes. Against $C_{4z}$ instability, they will move along the circle and finally annihilate at the edges $\mathcal{L}(\theta=0,\pi)$ or $\mathcal{L}(\theta=\frac{\pi}{2},\frac{3\pi}{2})$, as indicated by the black arrows. }
	\label{figS1}
\end{figure}

\subsection{Theory of edge states}
\label{edge theory}
\indent To understand the second-order topological superconductivity, we deduce an theory of edge states from the low-energy effective Hamiltonian, Eq.~(\ref{fullHeff}). We first notice that any edge termination of a 2D silce can be described as a tangent line of the unit circle geometry~\cite{33_from_Das,54_from_Das}, as shown in Fig.~\ref{figS1}(b). An arbitrary edge $\mathcal{L}(\theta)$ can be uniquely labeled by its unit normal vector defined as $\textbf{n}_{\mathcal{L}}=\left(\cos\theta,\sin\theta\right)^T$.  To solve edge dispersion for $\mathcal{L}(\theta)$, we apply Euler rotation around $z$ axis $R_Z(-\theta)$ to local coordinates, \ie $\bk'\equiv\left(k_1,k_2\right)^T=R_Z(-\theta)\left(k_x,k_y\right)^T$. Noticing that $k_1=\textbf{n}_{\mathcal{L}}\cdot\bkp$, now one can solve edge theory on arbitrary $\mathcal{L}(\theta)$ by imposing open boundary conditions on $H_{k_z\text{-plane}}(\bk')$ along the $k_1$ direction. However, $H_{k_z\text{-plane}}(\bk')$ generally has a complicated form in $\bk'$. Therefore, after the local coordinate rotation, we simultaneously apply a unitary transformation $\tilde{U}(\theta)=e^{-i\omega_{\Delta}\tilde{\Gamma}_{23}/2}e^{-i\theta\tilde{\Gamma}_{23}/2}$ on $H_{k_z\text{-plane}}(\bk')$ such that $\tilde{H}_{k_z\text{-plane}}(\bk',\theta)=\tilde{U}^\dagger(\theta) H_{k_z\text{-plane}}(\bk') \tilde{U}(\theta)=h_0+h_1$ has a simple form $\left(\mathrm{to}\ \mathcal{O}(k_1^2,k_2)\right)$,
	\begin{equation}\begin{aligned}
	&h_0(k_1)=
	(-M+Tk_1^2)\tilde{\Gamma}'_1
	-\vert\Delta\vert k_1\tilde{\Gamma}'_2,\\
	&h_1(\bk',\theta)=-\vert\Delta\vert k_2\tilde{\Gamma}'_3 - m(\theta)k_1^2\tilde{\Gamma}'_4.
	\end{aligned}\end{equation}
\noindent Here, $\vert\Delta\vert e^{i\omega_{\Delta}}$ $\equiv$ $\eta(\Delta_2+i\Delta_3)/\tilde{t_I}$, $m(\theta)$=$\gamma_1\cos{2\theta}+\gamma_2\sin{2\theta}$. Now we solve for the zero mode equation of $h_0$ and consider $h_1$ as a perturbation to extract the edge state dispersion of $\mathcal{L}(\theta)$.
\\
\indent Replacing $k_1 \rightarrow -i\partial_{x_1}$ yields the zero mode equation $h_0(-i\partial_{x_1})\psi(x_1)=0$. With the boundary conditions $\psi(x_1$=$0)=\psi(x_1$=$-\infty)=0$, two exponentially localized solutions $\psi_{1,2}$ are found near $x_1$=$0$:
	\begin{equation}\begin{aligned}
	\psi_i=\mathcal{N} \sin\left(\lambda_2 x_1\right) e^{\lambda_1 x_1} e^{ik_2 x_2} \phi_i, \quad i=1,2,
	\end{aligned}\end{equation}
\noindent where the eigen constants are
	\begin{equation}\begin{aligned}
	\lambda_1=-\frac{\vert\Delta\vert}{2T}, \quad \lambda_2=\frac{\sqrt{4MT-\vert\Delta\vert^2}}{2T},
	\end{aligned}\end{equation}
\noindent with $\mathcal{N}=2\sqrt{-M\vert\Delta\vert/(4MT-\vert\Delta\vert^2)}$ the normalization factor. The spinor part $\phi_i$ are eigenstates of $\Gamma_{12}$ with eigenvalue $+1$:
	\begin{equation}\begin{aligned}
	\phi_1=(\begin{array}{cccc}0&i&0&1\end{array})^T, \quad
	\phi_2=(\begin{array}{cccc}-i&0&1&0\end{array})^T.
	\end{aligned}\end{equation}
\noindent Treating $h_1$ as a perturbation, the surperconducting edge dispersion for any boundary $\mathcal{L}(\theta)$ is described by $\left(H_{\mathcal{L}}\right)_{ij}=\int_{-\infty}^0d{x_1}\int_{-\infty}^{+\infty}dx_2\psi_i^\dagger h_1 \psi_j$, thus
	\begin{equation}\begin{aligned}
	&H_{\theta}^{edge}(k_2) = \vert\Delta\vert k_2\bar{\sigma}_z + m_{\mathrm{eff}}(\theta)\bar{\sigma}_y,
	\end{aligned}\end{equation}
	\begin{equation}\label{meff}\begin{aligned}
	&m_{\mathrm{eff}}(\theta) = \frac{2(\tilde{t}_{\II}-\tilde{t}_{\I})(\tilde{\alpha}\Delta_2\cos{2\theta}+\tilde{\beta}\Delta_3\sin{2\theta})}{(\tilde{t}_{\II}+\tilde{t}_{\I})(t_a-\eta^2/\tilde{t}_{\I})}.
	\end{aligned}\end{equation}
\noindent $\bar{\bm{\sigma}}$ is in the sapce spanned by $\phi_{1,2}$. The fourfold rotation operator on the edge is $C_{4z}^{edge}(\theta)=-i\bar{\sigma}_z$. As a check, it acts on the matrix part of the edge mass term as $C_{4z}^{edge}(\theta)\bar{\sigma}_y	C_{4z}^{edge \dagger}(\theta)=-\bar{\sigma}_y$, and thus for the scalar part there is $m_{\mathrm{eff}}(\theta+\pi/2)=-m_{\mathrm{eff}}(\theta)$. Therefore, the sign flip of the edge mass term between $\mathcal{L}(\theta)$ and $\mathcal{L}(\theta+\pi/2)$ is protected by the fourfold rotation.
\\
\indent To consider the stability of second-order Majorana zero modes, the phenomenologically $C_{4z}$-breaking term $\delta\sin{k_z}\Gamma_4$ adding to $H_0(\bk)$ (Eq.~(\ref{HinS})) endows a constant tensor $m_\delta\bar{\sigma}_y$ to the edge dispersion $H_{\theta}^{edge}(k_2)$, which explicitly is that
\begin{equation}\label{mdelta}\begin{aligned}
	m_\delta=2\delta\Delta_2\frac{\eta^2(\tilde{t}_{\I}^2-\tilde{t}_{\II}^2)-2\tilde{t}_{\I}^2(t_a\tilde{t}_{\I}-\eta^2)}{\tilde{t}_{\I}^2(\tilde{t}_{\I}+\tilde{t}_{\II})^2(t_a-\eta^2/\tilde{t}_{\I})}\sin{k_z}.
\end{aligned}\end{equation}
\noindent Coupling to the anisotropic mass term $m_{\mathrm{eff}}(\theta)$, $m_\delta$ endows an extra constant gap for any edge $\mathcal{L}(\theta)$, which pushes the second-order Majorana zero modes to annihilate each other at the edges $\mathcal{L}(\theta=0,\pi)$ or $\mathcal{L}(\theta=\frac{\pi}{2},\frac{3\pi}{2})$.

\subsection{Topological characterization with nested Wilson loop}
\label{WL}

\begin{figure}[t!]
	\centering
	\includegraphics[scale=1.1]{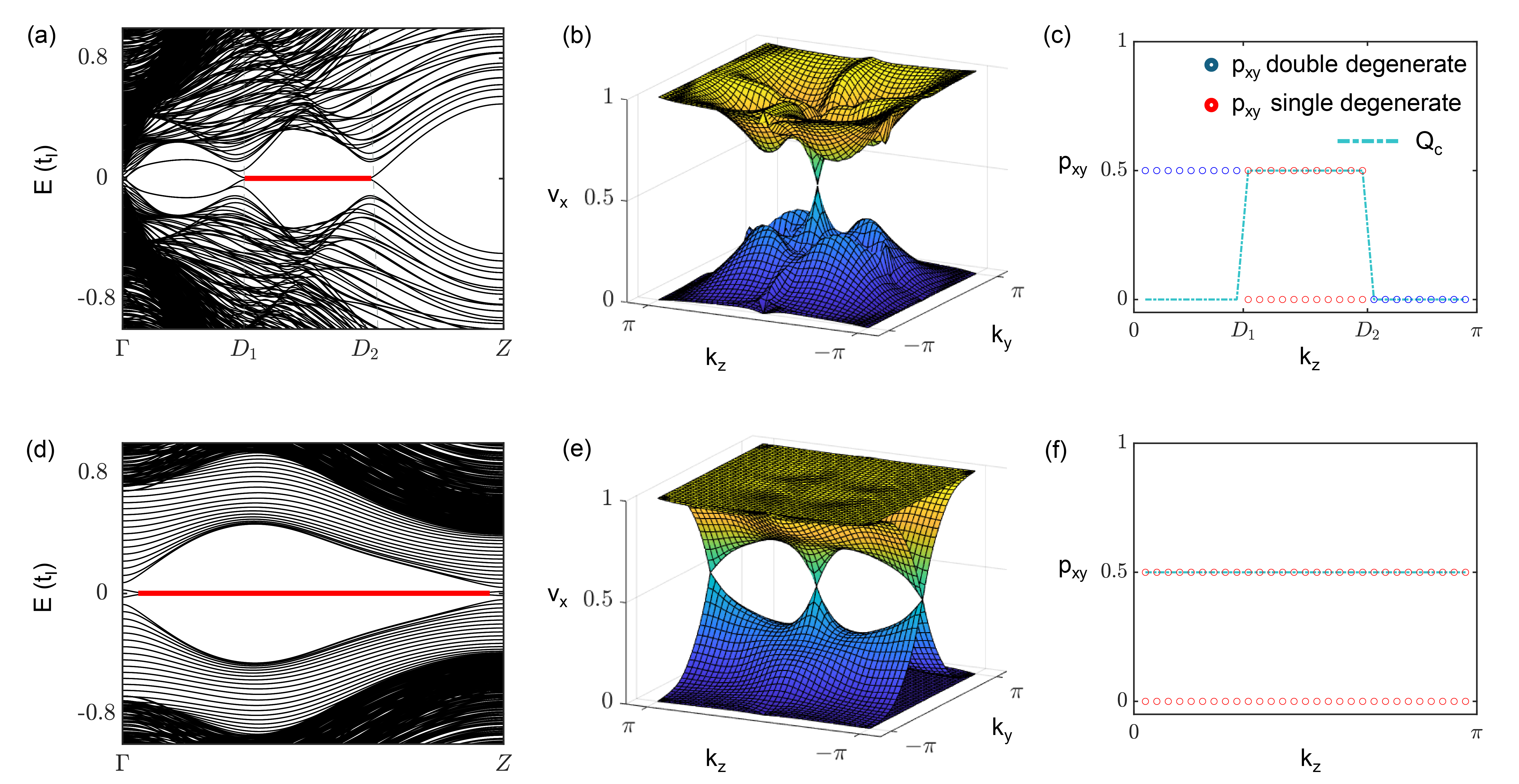}
	\caption{(a,~d) Energy dispersion in a wire geometry along $k_z$ with open boundary along both $x$ and $y$. (b,~e) Eigenvalues of Wilson loop operator of all occupied BdG bands of $H_{BdG}(\bk)$ (Eq.~(1) in the main text) along $k_x$. (c,~f) Dots: Eigenvalues of nested Wilson loop operator of all occupied wannier bands in (b,~e) along $k_y$. Lines: corner charge $Q_c$. $D_1$ and $D_2$ are the location of Superconducting Dirac points. (a-c) are for type-$\I$ DSMs. (d-f) are for type-$\II$ DSMs. In the nested Wilson loop calculation, the gapless $k_z$-planes in (b,~e) have been taken out. Note that here we have set the $B_{2u}$ pairing channel to be zero, $\Delta_3=0$.}
	\label{figS2}
\end{figure}

In this section, we discuss the topological characterization of the hinge Majorana states in our model using the concept of nested Wilson loop~\cite{quadrupole}. A full BdG Hamiltonian is diagonalized,
\begin{equation}
    H_{BdG}(\bk) = \sum_{\mu,\bk}\gamma_{\mu \bk}^\dagger E_{\mu \bk}\gamma_{\mu \bk},
\end{equation}
where $\gamma_{\mu \bk}^\dagger=\sum_\alpha u_{\mu \bk}^\alpha c_{\alpha \bk}^\dagger$, $\alpha=1,2,...,N_{\mathrm{orb}}$. $N_{\mathrm{orb}}$ is the number of orbitals. To calculate the Wilson loop along $k_x$ direction, we can construct the overlap matrix $F_x(k_y,k_z)$ using the occupied bands,
\begin{equation}
    F_{x,k_x^j}^{mn}(k_y,k_z)=\sum_\alpha u^{\alpha*}_{m,\bk^j} u^\alpha_{n,\bk^{j+1}}
\end{equation}
where $k_{x}^{j} = 2{\pi}j/N_x$ and $m, n = 1, 2, ... , N_{\rm occ}$, with $N_{\rm occ}$ is the number of occupied bands. Define $\bk'\equiv(k_y,k_z)$, hence,
\begin{equation}
    \mathcal{W}_{x}(\bk') = \prod_{j = 0}^{N - 1} F_{x,k_x^j}(\bk').
\end{equation}
Extracting the phase part, define a Wannier Hamiltonian $\mathcal{W}_{x}(\bk') = e^{iH_{\mathcal{W}_x}(\bk')}$ and diagonalize it, $H_{\mathcal{W}_x}(\bk')=2\pi \sum_{n=1}^{N_\mathrm{occ}}\xi^\dagger_{n,\bk'}\nu_x^n(\bk')\xi_{n,\bk'}$, with $\xi^\dagger_n(\bk')=\sum_{m=1}^{N_\mathrm{occ}}\tilde{u}^m_{n,\bk'}\gamma^\dagger_{m,\bk'}$. In the following Wannier band subsapces, $\xi^\dagger_n(\bk')=\sum_{\alpha=1}^{N_{\mathrm{orb}}}w^\alpha_{n\bk'}c^\dagger_{\alpha\bk'}$, with components $w^\alpha_{n\bk'}=\sum_{m=1}^{N_\mathrm{{occ}}}\tilde{u}^m_{n,\bk'}u^\alpha_{m,\bk'}$, the nested Wilson loop is similarly defined by constructed 
\begin{equation}
	\tilde{F}_{xy,k_y^j}^{mn}(k_z)=\sum_\alpha w^{\alpha*}_{m,\bk'^j} w^\alpha_{n,\bk'^{j+1}}
\end{equation}
where $k_{y}^{j} = 2{\pi}j/N_y$ and $m, n = 1, 2, ... , N_{\rm P}$. In our case, $N_{\rm P}=2$. And
\begin{equation}
    \mathcal{W}_{xy}(k_z) = \prod_{j = 0}^{N_y - 1} \tilde{F}_{xy,k_y^j}(k_z).
\end{equation}
Thus the final Berry phase of nested Wilson loop $p_{xy}$ is obtained by diagonalize $\mathcal{W}_{xy}(k_z)=\sum_{n=1}^{N_{{\rm P}}}\rho^\dagger_{n,k_z}e^{i2\pi p_{xy}^{n}(k_z)}\rho_{n,k_z}$. From nested Wilson loop, the equivalent ``corner charge" can be characterized as $Q_c=\sum_n p_{xy}^n\ {\rm mod}\ 1$.

In Fig.~\ref{figS2}, we have plotted the hinge states, eigenvalue of (nested) Wilson loop and corner charge for type-$\I$ (a-c) and type-$\II$ (d-f) DSMs.

In type-$\I$ DSMs, It is shown that the hinge Majorana states appear, linking the projection of two bulk SC DPs ($\bk=(0,0,D_{1,2})$). Accordingly, taking out two SC DPs, the eigenvalue of Wilson loop $v_x$ in $k_y$-$k_z$ plane shows a full gap, except $(k_y,k_z)=(0,0)$ point, at which the pumping indicate a set of mirror-symmetry protected Majorana surface modes discussed in Ref.~\cite{Sato_TCSC}. Meanwhile, the corner charge is characterized by the nested Wilson loop, showing the nontrivial Wannier charge center $Q_c=0.5$ between two bulk SC DPs.

In type-$\II$ DSMs, on the contrary, the bulk SC DPs are annihilating each other and creating a full bulk gap. Therefore, the intact hinge Majorana flat bands extends across the whole hinge Brillouin zone, linking the projection of the surface helical Majorana states. Correspondingly, the eigenvalue of Wilson loop $v_x$ shows $\mathbb{Z}_2$ pumpings at $(k_y,k_z)=(0,0/\pi)$ and is gapped out elsewhere. Meanwhile, taking out two ill-defined points ($k_z=0,\pi$), each $k_z$-plane shows an nontrivial corner charge $Q_c=0.5$, consisting to the hinge Majorana states.

\clearpage

\end{widetext}

\end{document}